\documentclass{article}
\usepackage{MnSymbol}

\usepackage{newlfont}
\usepackage{amsmath}
\usepackage{graphicx}
\usepackage{setspace}
\usepackage{cases}
\usepackage{stmaryrd}
\usepackage{epstopdf} 
\usepackage{mathrsfs}

\newcommand{\Og}{\mathcal{O}}

\begin{document}
%
%
%
%
\title{\textbf{The inversion of motion of bristle bots: analytical and experimental analysis}}
%
%
\author{%
    Giancarlo Cicconofri{*},
    Felix Becker{\dag},
    Giovanni Noselli{*}, 
		\\Antonio DeSimone{*} and
		Klaus Zimmermann{\dag} \\
  {*}SISSA, International School for Advanced Studies, Trieste, Italy
    \\ 
    {\dag}Technical Mechanics Group, Technische Universit\"{a}t Ilmenau, Germany
    }
%
%
\date{}
    \maketitle
%
%
%
%
    \noindent \begin{abstract}
		Bristle bots are vibration-driven robots actuated by the motion of an internal oscillating mass. Vibrations are translated into directed locomotion due to the alternating friction resistance between robots' bristles and the substrate during oscillations. Bristle bots are, in general, unidirectional locomotion systems. In this paper we demonstrate that motion direction of vertically vibrated bristle systems can be controlled by tuning the frequency of their oscillatory actuation. We report theoretical and experimental results obtained by studying an equivalent system, consisting of an inactive robot placed on a vertically vibrating substrate. 
%
    \end{abstract}
    %
%
%

\newpage

\tableofcontents

\newpage

    \section{Introduction}
Bristle bots are characterised by small size, robust and cheap design, and high speed of locomotion. Applications of bristle bots can be found in inspection technology \cite{Nr20}, search and rescue systems \cite{Nr19a}, and swarm robotic research \cite{Nr2}. The mechanism underlying their locomotion capabilities has been studied in          \cite{Nr18,Romansy,Nr2,GC}. To change motion direction of bristle-based mobile robots the following methods have been reported in the literature: changing the rotation direction of an unbalanced motor \cite{russian}, using the phase shift between two unbalanced rotors \cite{Nr18} or changing the inclination of the bristle system using additional actuators \cite{HartmannIWK}. Recent theoretical studies \cite{Nr1,GC} have suggested that, for systems excited by vertical oscillations and moving along a straight line, direction of motion can be controlled by tuning the frequency of actuation. We provide in this paper an experimental validation of this prediction. Our results may be of interest in the field of inspection systems optimized for limited manoeuvring space, e.g. pipe inspection robots \cite{Becker2014}.

The paper is organized as follows. In Section 2 we accommodate the analysis presented in \cite{GC} for internally actuated robots in the context of an equivalent system, which consists of an inactive robot placed on a vibrating substrate. This setting provides cleaner and more efficient experimental study. In Section 3 we summarize the results of the experiments, and in Section 4 we outline possible directions for future work. 
%
%
    \section{Setting, modelling, and analysis} 
\begin{figure}[htb!]
   \centering
   \includegraphics{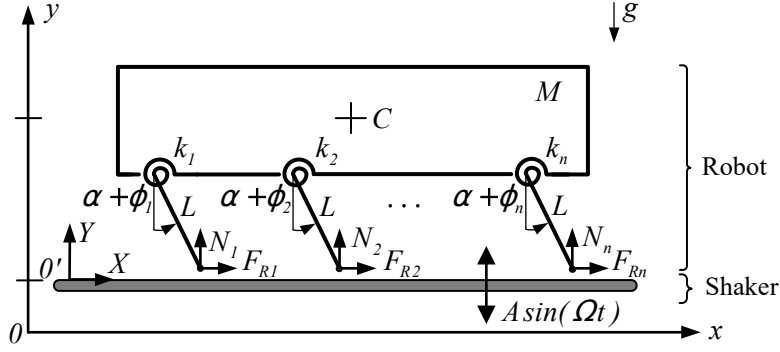}
   \caption{Schematic description of a model bristle bot lying on a shaker}
   \label{fig:ly-modell}
\end{figure}

		Bristle bots are actuated by an internal vibrating engine. In order to better study their behaviour experimentally, however, we can avoid the encumbrance of an on-board motor by considering the setting depicted in Fig.~\ref{fig:ly-modell}. The setting consists of a (inactive) robot lying on a vertically vibrating substrate (shaker). As we show below, the resulting physical system, when considered in the shaker attached frame, is identical to that of a bristle bot moving on a still substrate and driven by an internal oscillating force.  

The robot is modelled as a two-dimensional rigid object, consisting on a row of $m$ weightless support elements (bristles) of length $L$ attached to a main body of mass $M$. The $i$-th bristle is connected to the main body by a rotatory spring of stiffness $k_i$. The inclination of the bristles with respect to the vertical is given by $\alpha + \varphi_i$, where $\alpha$ is the inclination angle in the unloaded configuration. The (horizontal) friction force acting at the contact point of the $i$-th bristle with the shaker is modelled as
	\begin{equation}
		{F}_{Ri}=-\mu N_i \dot{{P}}_i \, ,
	\end{equation}
where $N_i$ is the normal reaction force acting on the tip of the bristle, $\mu$ is a phenomenological friction coefficient, and $\dot{P}_i$ the velocity of the contact point in the horizontal direction. We denote with a dot the derivative with respect to time.

%
%
%
    \subsection{Simplifying assumptions and equations of motion.} We introduce two Cartesian coordinate systems in the vertical plane: the fixed reference frame ($0xy$) and the shaker-attached frame ($0'XY$). The vertical displacement of the shaker at time $t$ with respect to the $x$ axis is given by $A \sin(\Omega t)$. We suppose that each bristle is always in contact with the shaker, and that the robot does not rotate with respect to the $xy$-plane. We have then 
		\begin{equation}		
		\varphi=\varphi_i=\varphi_1= ... =\varphi_m \, ,
		\label{winkel}
		\end{equation}
		while all contact points have the same horizontal velocity $\dot{P}_i=\dot{{P}}=\dot{x}+\dot{\varphi}L\cos(\alpha + \varphi)$ for every $i=1, \ldots , m$. Applying the principle of linear momentum we obtain
	\begin{equation*}
		M\ddot{y}=N-Mg \, , \quad M\ddot{x}=-\mu N \dot{P} \, ,
	\end{equation*}
where $x$ and $y$ are the coordinate of the centre of mass $C$ in the fixed frame, and $N=\sum\limits_{i=1}^m N_i$ is the total normal force. With $x=X$ and $y=Y+A\sin(\Omega t)$ the balance of linear momentum reads
	\begin{equation}
		M\ddot{Y}=N-Mg+MA\Omega^2\sin(\Omega t) \, ,
		\quad 
		M\ddot{X}=-\mu N \dot{P}.
		\label{onetwo}
	\end{equation}
	Notice that $X$ and $Y$ are the coordinates of $C$ in the shaker-attached frame. Finally, the principle of angular momentum gives
		\begin{equation}
		k \, \varphi  =  NL\sin(\alpha + \varphi) - \mu N\dot{P}L\cos(\alpha + \varphi),
		\label{three}
	\end{equation}
where $k=\sum\limits_{i=1}^m k_i$. Observe that equations (\ref{onetwo}) and (\ref{three}) are formally identical to the equations describing the same bristle bot model lying on a still substrate and actuated by an internal vertical force $F(t)= MA\Omega^2\sin(\Omega t)$.
%
%
%
\subsection{Nondimensionalization and order of magnitude of parameters.} 
To normalize the dynamical variables we define the following parameters 
\begin{equation*}
		\sigma = \sin(\alpha), \quad  \chi = \cos(\alpha) \quad and \quad \epsilon = \frac{M g L \, \sigma}{k} \label{epsilon} \, .
\end{equation*}
We define then the \emph{normalized} normal force $n$, angle difference $\theta$, and horizontal velocity $w$ of the robot as
\begin{equation*}
 n=\frac{N}{M g } \, , \quad \theta = \frac{\varphi}{\epsilon} \, , \quad \textrm{and} \quad w = \frac{\dot{X}}{\epsilon \Omega L \chi}  \,.  \label{n-theta-w}
\end{equation*}
Applying all the definitions above, equations (\ref{onetwo}) and (\ref{three}) can be rewritten as the equivalent system in the dimensionless time $\tau=\Omega t$
\begin{equation}
\left\{
	\begin{aligned}
&	- \gamma \ddot{\theta} \, \frac{\sin (\alpha + \epsilon\theta)}{\sigma}  - \epsilon  \left(\frac{\chi\gamma}{\sigma}\right)\dot{\theta}^{2}\frac{ \cos(\alpha + \epsilon\theta)}{\chi} =  n - 1 + \eta \sin \tau  \\
&\dot{w} =-\lambda  \, n \left(w + \dot{\theta} \,\frac{\cos(\alpha  +\epsilon\theta)}{\chi} \right) \\
&\theta  = \: n \, \frac{\sin(\alpha + \epsilon \theta)}{\sigma} \: - \: \xi  \, n  \left( w + \dot{\theta} \,\frac{\cos(\alpha +\epsilon\theta)}{\chi} \right)\frac{\cos(\alpha +\epsilon\theta)}{\chi} 
\end{aligned} \right. \label{norm}
\end{equation}
where
\begin{equation}
		\xi  = \frac{\mu M g L^{2}\chi^{2}\Omega}{k} , \quad   \lambda = \frac{\mu g }{\Omega} \quad , \quad \gamma =  \frac{(L \sigma \Omega )^2 M}{k}  \quad \textrm{and} \quad \eta=\frac{ A M \Omega^{2}}{Mg} \, .  \label{xilambda}
\end{equation}
In the following we suppose that $\eta < 1$ is a small parameter, and $\epsilon  \lesssim \eta^{2}$.
%
%
\subsection{Asymptotic analysis and average velocity.} 
We derive in this section an estimate of the average horizontal velocity of the robot, and we show how it can change sign for different values of the frequency of actuation. To obtain this estimate, we solve (\ref{norm}) by expanding the solution in power series in the (small) parameter $\eta$, that is
\begin{equation}
\theta = \theta_{0} + \eta\theta_{1}  + \eta^2 \theta_{2} + \ldots\, , \quad   w = w_{0} + \eta w_{1} + \eta^2 w_{2} + \ldots ,  \quad  n =  n_{0} + \eta n_{1} + \eta^2 n_{2} + \ldots 
 \label{expa}
\end{equation}
Expanding (\ref{norm}) in powers of $\eta$, and matching coefficients of equal power, leads to a sequence of equations to be solved successively for the unknowns ($\theta_j, w_j, n_j$), with $j=1,2,...$. It can be proved rigorously, see \cite{GC}, that \eqref{expa} converge uniformly for every small enough $\eta$, and that only one periodic solution exist for each coefficient $\theta_i$, $w_{i}$ and $n_{i}$ at each order. The resulting sum \eqref{expa} for $\theta$, $w$, and $n$ is the only periodic solution of \eqref{norm}, and any other solution of the system converges asymptotically in time to it.

The zero-order system is given by 
\begin{displaymath}
	 - \gamma \ddot{\theta}_{0} = n_{0} -1 	\, , \quad  \dot{w}_{0}  = - \, \lambda \left( w_{0} + \dot{\theta}_{0}\right) \, , \quad \theta_{0}  = n_{0}  - \xi \left( w_{0} + \dot{\theta}_{0} \right) \, ,
\end{displaymath}
and its only periodic solution is 
\begin{equation}
		\theta_{0} = 1, \quad  w_{0} = 0 \quad \textrm{and} \quad n_{0}=1 . \label{zeroord}
\end{equation}
Now, imposing \eqref{zeroord}, the first order system reads
\begin{equation}
 - \gamma \ddot{\theta}_{1} =  n_{1} + \sin \tau \, , \quad  \dot{w}_{1}  =  -\lambda \left( w_{1} + \dot{\theta}_{1}  \right) \, , \quad
\theta_{1}  =  n_{1} -  \xi \left( w_{1} + \dot{\theta}_{1}  \right) \, . \label{firstord}
\end{equation}
We look here for solutions of the type 
\begin{equation}
\theta_1 = \theta_{1}^s \sin \tau + \theta_{1}^{c} \cos \tau , \: \:  w_1= w_{1}^s \sin \tau + w_{1}^{c} \cos \tau , \: \: n_1 =  n_{1}^s \sin \tau + n_{1}^{c} \cos \tau . \label{sincos}
\end{equation}
Replacing \eqref{sincos} in \eqref{firstord} and matching coefficients of sines and cosines respectively, we end up with six equations which allow us to determine $\theta_{1}^s$ ,$\theta_{1}^{c}$, $w_{1}^s$, $w_{1}^{c}$,  $n_{1}^s$, and $n_{1}^{c}$. We obtain
\begin{align*}
	\theta_{1}^{s} & = \frac{(\gamma-1)(1+\lambda^2)+\lambda \xi}{(\gamma -1)^2 + ((\gamma-1)\lambda + \xi)^2} \, ,  &  \theta_{1}^{c} & =  \frac{\xi}{(\gamma -1)^2 + ((\gamma-1)\lambda + \xi)^2} \, ,  \\
w_{1}^{s} & = \frac{-(\gamma-1)\lambda}{(\gamma -1)^2 + ((\gamma-1)\lambda + \xi)^2}  \, ,  &
w_{1}^{c} & = \frac{-(\gamma-1)\lambda^2 - \lambda\xi}{(\gamma -1)^2 + ((\gamma-1)\lambda + \xi)^2} \, ,  \\
n_{1}^{s} & =  \frac{(\gamma-1)(1+\lambda^2) - (\gamma - 2)\lambda  \xi - \xi^2}{(\gamma -1)^2 + ((\gamma-1)\lambda + \xi)^2} \, ,   &
n_{1}^{c} & =  \frac{\gamma \xi}{(\gamma -1)^2 + ((\gamma-1)\lambda + \xi)^2} \, .
\end{align*}   
We then recover, in particular, $w= \eta w_{1} + \Og(\eta^2)$, where $w_1$ is a periodic function with zero average. Indeed, the average velocity of the robot is of the order $\sim \eta^2$, however, we do not need to solve the second order system to recover a formula for it. We observe that, imposing \eqref{zeroord}, the second order expansion of the second equation in \eqref{norm} gives
\begin{equation}
	\dot{w}_{2} = -\lambda (w_2 +\dot{\theta}_2) - \lambda n_{1} (w_1 +\dot{\theta}_1) \label{secondord}\, .
\end{equation}
We know from the previously stated results in \cite{GC} that \eqref{secondord} admits one periodic solution for $w_2$ and $\theta_2$. Therefore, in particular, $\dot{w}_2$ and $\dot{\theta}_2$ have zero average. From \eqref{secondord} then follows that the average $w^*$ of $w_2$ can be written in terms of the solution of the first order system
\begin{equation}
w^* := \frac{1}{2 \pi} \int_{0}^{2 \pi} \! \!  w_2  = \frac{- 1}{2 \pi} \int_{0}^{2 \pi} \! \!  n_{1} (w_1 +\dot{\theta}_1)  = -\frac{1}{2}\left(\frac{\xi - \lambda}{(\gamma -1)^2 + ((\gamma-1)\lambda + \xi)^2} \right) \, . \label{meanv}
\end{equation}
This last equation provides an explicit formula for the approximate (normalized) average horizontal velocity of the robot since
\begin{displaymath}
	\frac{1}{2 \pi} \int_{0}^{2 \pi} \! \! w = \eta^2  w^* + \Og(\eta^3) \, .
\end{displaymath}
Moreover, \eqref{meanv} shows how the sign of the average velocity depends on that of the difference between the two parameters $\xi$ and $\lambda$, and ultimately on the frequency $f:=\Omega/2\pi$, see \eqref{xilambda}.  The formula predicts an average motion in the negative direction for large values of $f$, and in the positive direction for small values of $f$. Fig. 2 shows the frequency dependence of $w^*$ when we fit \eqref{meanv} with the parameters of the prototype described below. The frequency such that $w^*=0$ is given by
\begin{equation}
	f_{\textrm{inv}} = \frac{1}{2\pi} \cdot \frac{\sqrt{k/M}}{ L \cos(\alpha)}  \, ,
	\label{inver}
\end{equation}
which gives an approximation of the frequency at which the inversion of motion of the robot occurs. In the experiments below $f_{\textrm{inv}} \simeq 14 \textrm{Hz}$.
%
%



%
%
%
%
%
\begin{figure}[h!]
   \centering
   \includegraphics{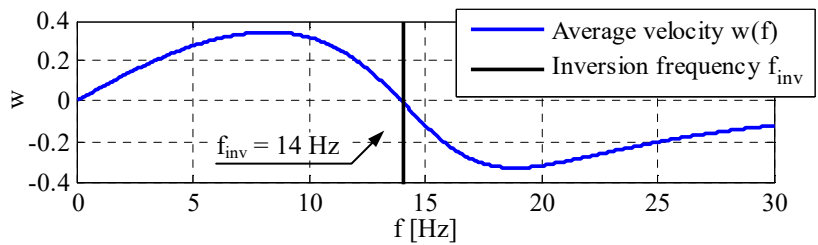}
   \caption{Average velocity $w^*$ against excitation frequency $f$, with $M = $10.5~g, $L =$ 8~mm, $\mu = 5$, $k = 3.5\cdot 10^{-3}$~Nm, $g~=$~9.81~$\frac{m}{s^2}$, $\alpha = 35.2^\circ$}
   \label{simulation}
\end{figure}
%
%
\section{Experiments} 
\subsection{Setup.} 
\begin{figure}[htb!]
   \centering
   \includegraphics{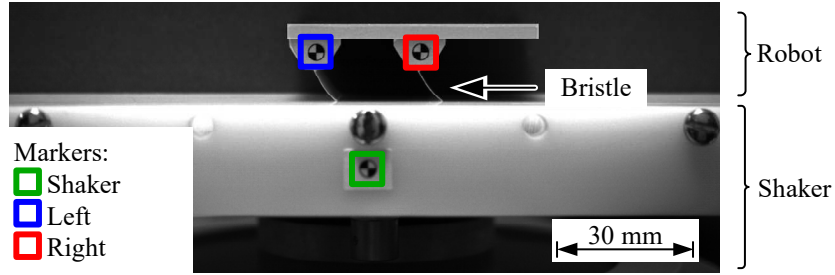}
   \caption{Experimental setup (single frame of a slow motion video)}
   \label{fig:exp}
\end{figure}

The experimental setup is shown in Fig.~\ref{fig:exp}. It consists of a passive robot prototype lying on a platform attached to an electromagnetic shaker, which provides vertical excitation. The main body of the robot is made of polymer material with length $\times$ width $\times$ height = 55~mm $\times$ 35~mm $\times$ 17~mm., and mass $M=$~10.5~g. The bristle functionality is realised by two 30~mm wide paper strips with a free length $L=$~8~mm. With a mass of 55~mg, the paper strips meet sufficiently well the model assumption of massless bristles. The centre of mass of the robot is located in the middle between the ground-bristle contact points in order to avoid rotation on the main body, see model condition (\ref{winkel}). In contrast with the model, the elasticity of the real bristles is equally distributed along their length. Their equivalent rotational stiffness and inclination angle are calculated to be $k =~3.5\cdot 10^{-3}$~Nm and $\alpha=$~35.2~$^\circ$.  Robot and shaker are equipped with markers for motion tracking.

%

%


\subsection{Experimental procedure and results} The shaker is switched on producing vertical sinusoidal vibrations with controllable frequency and amplitude, leading to directed locomotion of the robot. At different frequencies we tune the amplitude of the shaker in order to match our analytic assumption $\eta<1$ and, in turn, to avoid the robot from losing contact with the ground. We recover a clear motion in the positive horizontal direction for frequencies below $10$ Hz, and motion in the negative direction for frequencies above $18$ Hz, in agreement with the theoretical predictions (between 10 and 18 Hz results are inconclusive).

\begin{figure}[h!]
   \centering
   \includegraphics{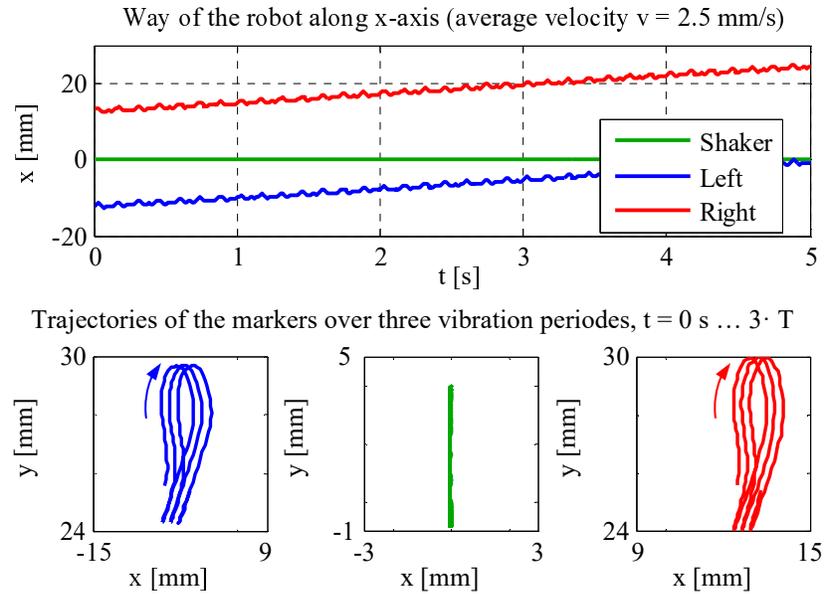}
   \caption{Locomotion behaviour of the prototype excited at 7~Hz}
   \label{fig:tracking1}
\end{figure}
\begin{figure}[h!]
   \centering
   \includegraphics{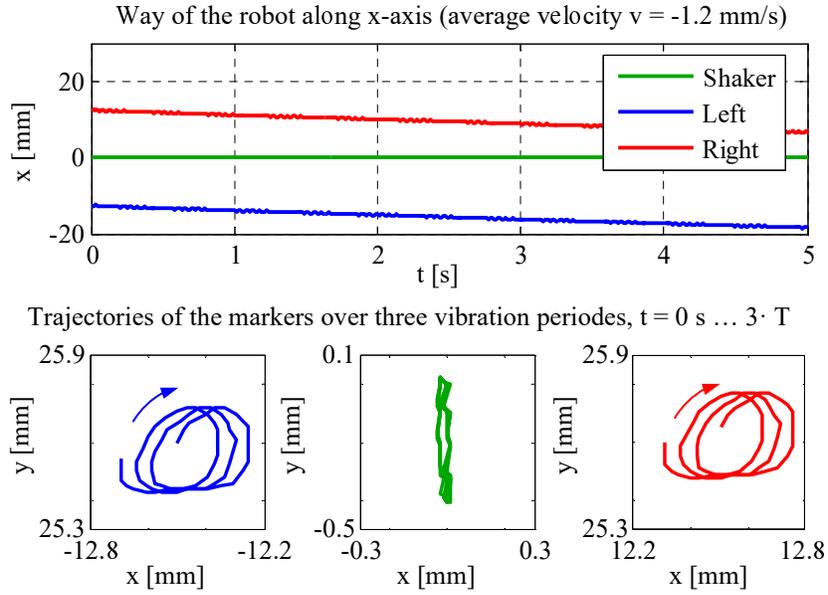}
   \caption{Locomotion behaviour of the prototype excited at 21~Hz}
   \label{fig:tracking2}
\end{figure}

\subsection{Two exemplarily locomotion characteristics} For two oscillation frequencies we filmed the experiments with an high-speed camera (Fig 3 shows a frame of the videos). Locomotion is analysed by tracking the markers on the robot and the shaker. Fig 4 presents the tracking results for an excitation below the calculated inversion frequency, while Fig. 5 shows the tracking results for excitation above the inversion frequency.

\section{Conclusions and outlook} 
We showed analytically and experimentally that the inversion of motion of bristle bots is possible by tuning the frequency of pure vertical excitation. Future work should focus on models accounting on more quantitatively accurate description of frictional interactions. Further experimental analysis is needed to find precisely the relation between robot parameters and locomotion characteristics.


\newpage


\begin{thebibliography}{25}

\bibitem{Romansy} F. Becker, V. Lysenko, V. Minchenya, I. Zeidis, and K. Zimmermann. An approach
to the dynamics of a vibration-driven robot. In \emph{Proc. of Romansy 19}, pages 299–308, 2013.

\bibitem{Becker2014} F. Becker, S. B\"{o}rner, T. B\"{o}rner, V. Lysenko, I. Zeidis, and K. Zimmermann. Spy
bristle bot a vibration-driven robot for the inspection of pipelines. In \emph{Proc.
58th IWK}, 2014.

\bibitem{GC} G. Cicconofri and A. DeSimone. Motility of a model bristle-bot: A theoretical
analysis.\emph{ International Journal of Non-Linear Mechanics}, 76:233–239, 2015.


\bibitem{Nr1} A. DeSimone and A. Tatone. Crawling motility through the analysis of model
locomotors: Two case studies. \emph{Eur. Phys. J. E.}, 35(85), 2012.


\bibitem{Nr2} L. Giomi, N. Hawley-Weld, and L. Mahadevan. Swarming, swirling and stasis in
sequestered bristle-bots. \emph{Proc. R. Soc. A}, 469, 2013.



\bibitem{Nr19a} K. Hatazaki, M. Konyo, K. Isaki, S. Tadokoro, and F. Takemura. Active scope
camera for urban search and rescue. In \emph{IEEE IROS}, pages 2596–2602, 2007.


\bibitem{Nr18} K. Ioi. A mobile micro-robot using centrifugal forces. In \emph{Proc. on Int. Conf. on
Advanced Intelligent Mechatronics}, pages 736–741, 1999.


\bibitem{HartmannIWK} M. Schulke, L. Hartmann, and C. Behn. Worm-like locomotion systems: devel-
opment of drives and selective anisotropic friction structures. In \emph{Proc. of 56th
IWK}, 2011.



\bibitem{russian} A. Senyutkin. Bristle bot. \emph{Young Technician}, 6:65–67, 1977. http://zhurnalko.net/
=sam/junyj-tehnik/1977-06–num53 (in Russian), Last visited: 14.12.2015.



\bibitem{Nr20} Z. Wang and H. Gu. A bristle-based pipeline robot for ill-constraint pipes. \emph{IEEE
Trans. Mechatron.}, 13(3):383–392, 2008.


\end{thebibliography}
\end{document}